\documentclass[fleqn,10pt]{wlscirep}
\usepackage[utf8]{inputenc}
\usepackage[T1]{fontenc}
\usepackage{lineno}
\usepackage{float}
\title{Brazilian COVID-19 data streaming}
\author[1,$\dag$,$\ast$]{Nívea B. da Silva}
\author[2,$\dag$]{Luis Iván O. Valencia}
\author[3]{Fábio M. H. S. Filho}
\author[2]{Andressa C. S. Ferreira}
\author[2]{Felipe A. C. Pereira}
\author[2,5]{Guilherme L. de Oliveira}
\author[2]{Paloma F. Oliveira}
\author[3]{Moreno S. Rodrigues} % Orcid 0000-0002-1594-2311
\author[2]{Pablo I. P. Ramos}
\author[2,4,$\dag$,$\ast$]{Juliane F. Oliveira}
 
\affil[1] {Federal University of Bahia, Department of Statistics, Salvador, Bahia,  40170-110, Brazil.}
\affil[2]{Center for Data and Knowledge Integration for Health (CIDACS), Instituto Gon\c calo Moniz, Funda\c c\~ao Oswaldo Cruz\\ Bahia, 40196-710, Brazil.}
\affil[3]{Fundação Oswaldo Cruz - Rondônia, Porto Velho, Rondônia, 76812-245, Brazil}
\affil[4]{Centre of Mathematics of the University of Porto (CMUP), Department of Mathematics, Porto, 4169-007, Portugal.}
\affil[5]{Federal Center for Technological Education of Minas Gerais, Belo Horizonte, Minas Gerais, 30510-000, Brazil.}

\affil[*]{corresponding author(s): Nívea Bispo (nivea.bispo@gmail.com) and Juliane Oliveira (julialanzin@gmail.com).}
\affil[$\dag$]{these authors contributed equally to this work.}

\bigskip

\begin{abstract} % não pode exceder 170 palavras %%%% 86 palavras
We collected individualized (unidentifiable) and aggregated openly available data from various sources related to suspected/confirmed SARS-CoV-2 infections, vaccinations, non-pharmaceutical government interventions, human mobility, and levels of population inequality in Brazil. In addition, a data structure allowing real-time data collection, curation, integration, and extract-transform-load processes for different objectives was developed. The granularity of this dataset (state- and municipality-wide) enables its application to individualized and ecological epidemiological studies, statistical, mathematical, and computational modeling, data visualization as well as the scientific dissemination of information on the COVID-19 pandemic in Brazil.
\end{abstract}

\begin{document}
\flushbottom
\maketitle

\thispagestyle{empty}

\section*{Background \& Summary} 

December 2021 marks two years since the first cases of SARS-CoV-2 were reported. For the first time in history, nearly real-time data on cases, deaths, vaccination administration, etc. has become available from multiple sources to describe the most important pandemic in the last century\cite{johns,worldometer21,whodash21}. While this information can help evaluate the dynamics of the pandemic, to better measure the impact of SARS-CoV-2 transmission, clinical and epidemiological information is required on patients suffering from COVID-19, as well as other aspects related to health systems and surveillance, human culture, behavior and mobility and social inequalities. More than ever, innovative data access processes and methods have become essential to understanding, managing and mitigating the effects of a pandemic in any region of the world\cite{world2021covid}.

Our objective was to provide users with essential data, updated in real-time, required to track the evolution of the COVID-19 pandemic in Brazil. Our approach compiled data available from open access platforms, including Google Mobility\cite{google}, the Brazilian Institute of Geography and Statistics (IBGE)\cite{ibge2016,ibge2019}, the Brazilian Unified Health System (SUS) OpenData platform\cite{ministerio2020SG, ministerio2021SRAG, ministerio2021vaccine}, The Platform for Analytical Models in Epidemiology (PAMEpi) via cooperation with legal information start-up JusBrasil\cite{PAMEpi2020,jus2022} and the Brazilian Deprivation Index (BDI, developed by CIDACS/Fiocruz and the University of Glasgow)\cite{allik2020developing,ibp}. The OpenData SUS platform provides individualised (unidentifiable) data on patients with suspected and confirmed SARS-CoV-2 infection, as well as information on populational coverage of the Brazilian vaccination campaign. Additionally, we have aggregated information on human mobility (using as data sources Google Mobility and IBGE), deprivation levels (BDI) and non-pharmaceutical interventions (PAMEpi - JusBrasil). 

\begin{figure}[H]
\centering
\includegraphics[height=8cm,width=16cm]{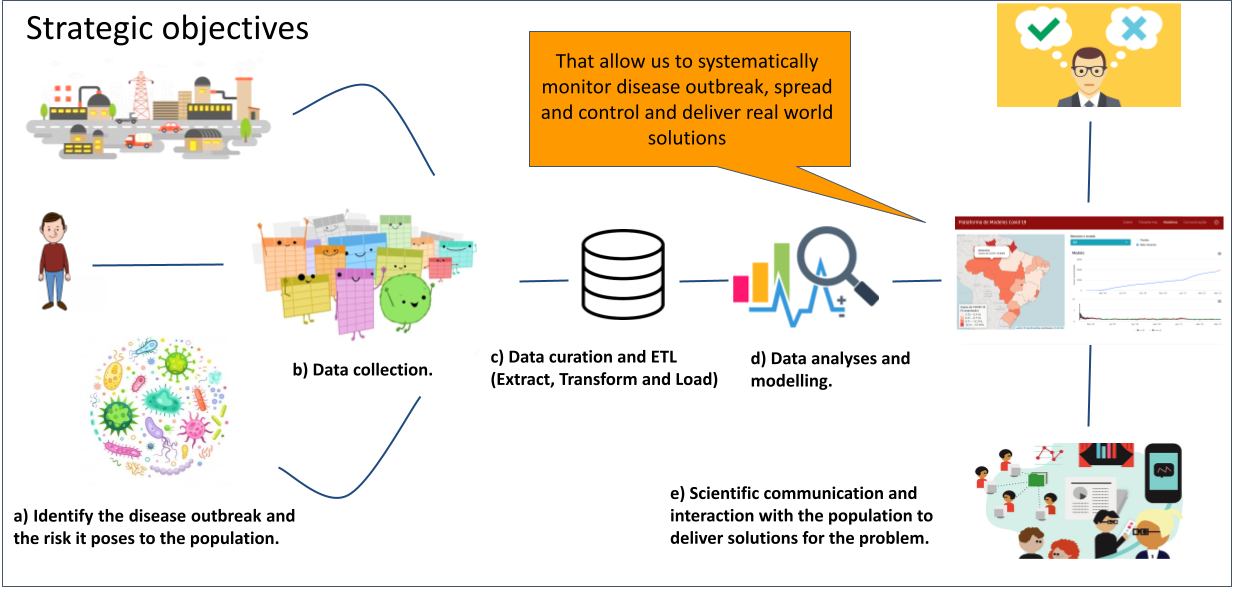}
\caption{The objectives of the Platform for Analytical Models in Epidemiology (PAMEpi).}
\label{pam}
\end{figure}

Motivated by the urgent need to support a rapid response to the COVID-19 pandemic, our team created The Platform for Analytical Models in Epidemiology (PAMEpi)\cite{PAMEpi2020}. This platform aims to provide a user-friendly interface, enabling researchers and policymakers to assess the impact of the spread of (emerging) infectious diseases in their respective areas of interest. To this end, the project has five main objectives: 1) Allow users to identify infectious disease threats; 2) Perform data collection that can provide users with the information needed to answer real-world questions; 3) Perform data curation and ETL (Extract, Transform, Load) processed; 4) Develop and apply data analysis and modelling scenarios; 5) Provide users with scientific information to communicate to stakeholders and the public in general (Figure \ref{pam}). By integrating these five objectives, we can harness the power of health data analytics to respond to COVID-19 and other disease outbreaks or future pandemics, as well as address health challenges, including conducting research on prevention, transmission and control.

Due to the continental size of Brazil, which ranks among the countries with the highest numbers of reported COVID-19 cases and deaths, the collection and curation of primary data on health, human mobility and non-pharmaceutical interventions, as well as information on the social and economic conditions of the population are challenging tasks. PAMEpi’s aims includes disease modeling, dissemination of scientific information on the COVID-19 pandemic at individual or aggregated levels and the promotion of multi-national cooperative studies, for which clean, harmonised and normalised data is a fundamental resource\cite{katikireddi2022two,wang2022estimating}. Considering the relevance of open science, the dataset we produced is essential to inform timely policy responses and aid in the understanding of the pandemic’s impact on other outcomes. In addition, this data architecture offers great potential for reuse by Brazilians and the international community, enabling comparative studies.

\bigskip

\section*{Methods}
\medskip

We collected individualized and aggregated openly available data from various sources related to suspected/confirmed SARS-CoV-2 infections, vaccinations, non-pharmaceutical government interventions, human mobility and deprivation index values.

All individualized and unidentifiable information related to suspected and confirmed SARS-CoV-2 infections and vaccinations (Vac\textit{db}) is available via the OpenData SUS system, provided by the Brazilian Ministry of Health (MoH)\cite{ministerio2020SG,ministerio2021SRAG,ministerio2021vaccine}. SARS-CoV-2 infections are notified via two databases: the Flu Syndrome Database (FS\textit{db}), which contains all suspected and confirmed mild-to-moderate COVID-19 cases, and the Severe Acute Respiratory Syndrome Database (SARS\textit{db}), containing cases of severe-to-critical infection. FS\textit{db}, SARS\textit{db} and Vac\textit{db} are licensed under a Creative Commons Attribution License (CC BY v4.0) and no ethical approval by an institutional review board was required to use this data (Brazilian National Health Council Resolutions 466/2012 and 510/2016, Article 1, Sections III and V). All data can be freely downloaded directly at the OpenData SUS website\cite{ministerio2020SG,ministerio2021vaccine,ministerio2021SRAG} or our Python algorithm can be used to download all data documented in this manuscript. It is important to mention that, in accordance with the Brazilian General Data Protection Law – 13.709/2018 (LGPD), some personally identifiable information has been suppressed from the datasets provided by the OpenData SUS portal, which reduces the total number of variables described in the respective data dictionaries.

With regard to aggregated data, Google Mobility enables users to access information on daily human mobility patterns at grocery stores and pharmacies, parks, transit stations, retail and recreation entities, residences and workplaces. In addition, historical series detailing average human flows on roads, rivers and via air are made freely available by the IBGE\cite{ibge2016,ibge2019}. Other collected aggregated information was provided through a cooperative agreement with legal information start-up JusBrasil\cite{jus2022}, which allowed automatic retrieval of Brazilian decrees describing non-pharmaceutical measures enacted by state or municipal governments to contain the pandemic. This database was obtained as raw text files and were subjected to human curation within PAMEpi, with enriched output made freely available as text and stringency values representing government containment efforts at state or municipal levels. Lastly, we also gathered information on community-level deprivation measures to assess socioeconomic inequalities.

\begin{figure}[H]
\centering
\includegraphics[height=10cm,width=17cm]{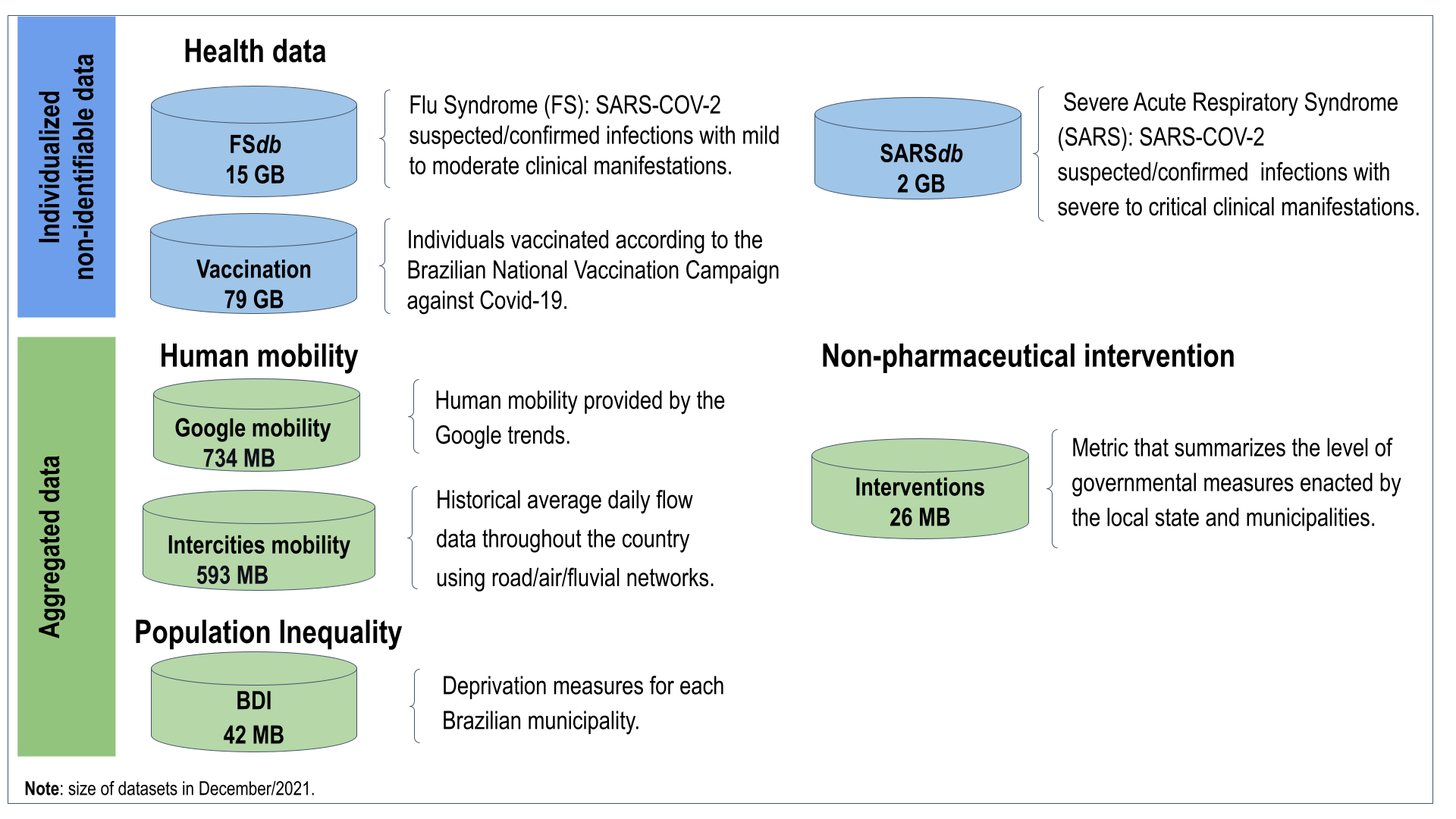}
\caption{Schematic overview of the Brazilian COVID-19 Data Lake.}
\label{fig1}
\end{figure}

Together, this collected data forms what we call the Brazilian COVID-19 Data Lake. Figure \ref{fig1} schematically illustrates the data sources and their respective characteristics.

For the creation of the data lake, all raw data was obtained in its original file format (usually CSV text files). Deprivation index and intercity mobility datasets are static, with updates expected only following census tabulation. Data from the other sources may be updated daily, weekly or monthly depending on the respective source. Scripts in Python language were elaborated to create and update the data lake automatically\cite{git2022}. Users can download and organize data locally, on an isolated partition or in an online repository, and can run these tasks manually or schedule them as system tasks. Due to the large volume of individualized information stored across several databases, an API available from the OpenData SUS portal can be used to update the information in the data lake.

The next step involves data curation. For this, we 1) examine field names, variable types and categorical values (if applicable) and compare them to previous versions of the data at each moment of data updating; 2) perform semantic enrichment, data harmonization and cleaning to facilitate the use of information from multiple databases; 3) normalize data, to better structure the dataset by reducing data redundancy and improving data integrity; 4) generate metadata. The environment used to load and process these datasets was built using Python (PySpark tool) and all codes to perform these tasks are available from PAMEpi GitHub code repository\cite{git2022}.

\begin{figure}[H]
\centering
\includegraphics[height=4cm,width=9cm]{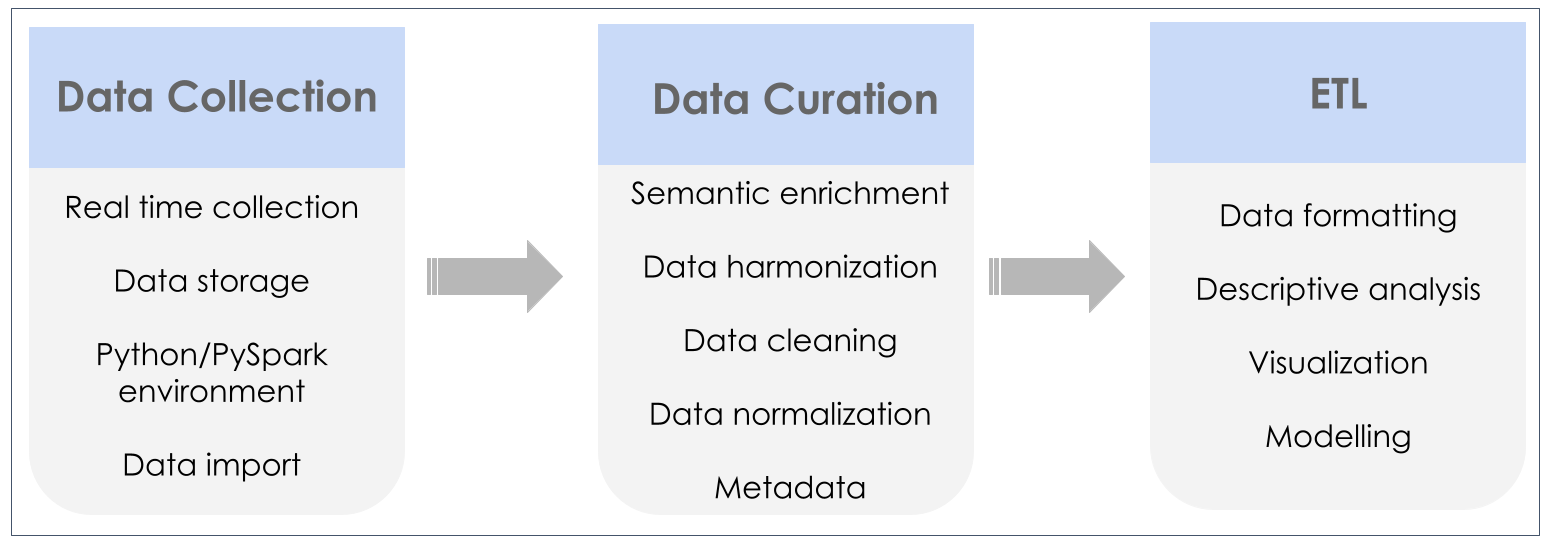}
\caption{Data lake framework for Brazilian COVID-19 big data architecture}
\label{datalake}
\end{figure}

The last stage involves formatting data for analysis and modelling. Both individualised and aggregated data analyses can be performed according to the goals of each researcher. As standard practice, we generate datasets aggregated at state, health region and municipal levels throughout Brazil. This comprises one of the most important outputs of our data architecture, which facilitates epidemiological studies, data visualization, health surveillance, etc. Figure \ref{datalake} details the data collection, curation and ETL operations of the PAMEpi data architecture. 

\section*{Data Records}

In the following two sections, we describe key information about the datasets that comprise the Brazilian COVID-19 Data Lake. Although the PAMEpi data architecture utilizes a limited number of datasets, the structure described is adaptable and can integrate any other datasets needed to address public health challenges.

All generated metadata and dataset descriptions are freely available via our online platform (\url{https://pamepi.rondonia.fiocruz.br/en/data_en.html}). To facilitate visualisation, we have provided a data explorer that allows users to view the first rows of each dataset along with metadata, including column descriptions, variable type, and variable harmonisation where applicable. This also allows broader re-use of this dataset, particularly since the original descriptors and data dictionaries are usually only available in Portuguese.

\subsection*{Individualized COVID-19 data}

\medskip

\subsubsection*{1. Flu Syndrome (FS) and Severe Acute Respiratory Syndrome (SARS) databases} %

\medskip

The advent of the COVID-19 pandemic in March 2020 prompted the MoH to implement a scheme to report mild-to-moderate suspected cases of COVID-19, denominated the Flu Syndrome database (FS\textit{db}). The MoH characterizes a patient as suspected of having influenza syndrome (IS) if at least two of the following symptoms are present: fever (measured or reported), chills, sore throat, headache, cough, runny nose, olfactory or taste abnormalities. In children, nasal congestion is another condition warranting consideration. In elderly individuals, syncope, mental confusion, excessive drowsiness, irritability, and loss of appetite may also constitute relevant symptoms\cite{ministerio2020guia}.

The Severe Acute Respiratory Syndrome (SARS) database (SARS\textit{db}) was created by the MoH through the Health Surveillance Secretariat after the last influenza virus subtype A (H1N1) pandemic in 2009. This database contains reports of influenza and other respiratory viruses, which were previously recorded only via influenza syndrome sentinel surveillance. After March 2020, severe-to-critical COVID-19 cases also began to be integrated into this database. According to the MoH, cases are reported in SARS\textit{db} if one of the following symptoms is present in addition to two other IS symptoms: shortness of breath/breathing difficulties, persistent chest discomfort or pain, O$_2$ saturation <95\%, or bluish discolouration (cyanosis) of the lips or face. In addition, tachypnea, hypoxemia, cyanosis, intercostal retraction, dehydration, and loss of appetite are relevant symptoms among children. In the elderly, anosmia, ageusia, diarrhea, abdominal pain, myalgia or symptoms of exhaustion may warrant consideration in determining inclusion in SARS\textit{db}\cite{ministerio2020guia}. Importantly, all cases reported in SARS\textit{db} required hospitalization due to disease severity.

A COVID-19 case can be diagnosed via three methods: clinical-epidemiological investigation, laboratory testing or by imaging. Clinical-epidemiological diagnosis is based on a patient’s clinical manifestations and possible contact with other infected individuals 14 days prior to the onset of symptoms; laboratory diagnosis can be confirmed by molecular testing (real-time PCR), serology (ELISA, CLIA or ECLIA) or by rapid tests; imaging by high resolution computed tomography can also provide evidence of COVID-19 disease\cite{ministerio2020guia}.

\begin{figure}[H]
\centering
\includegraphics[height=9cm,width=15cm]{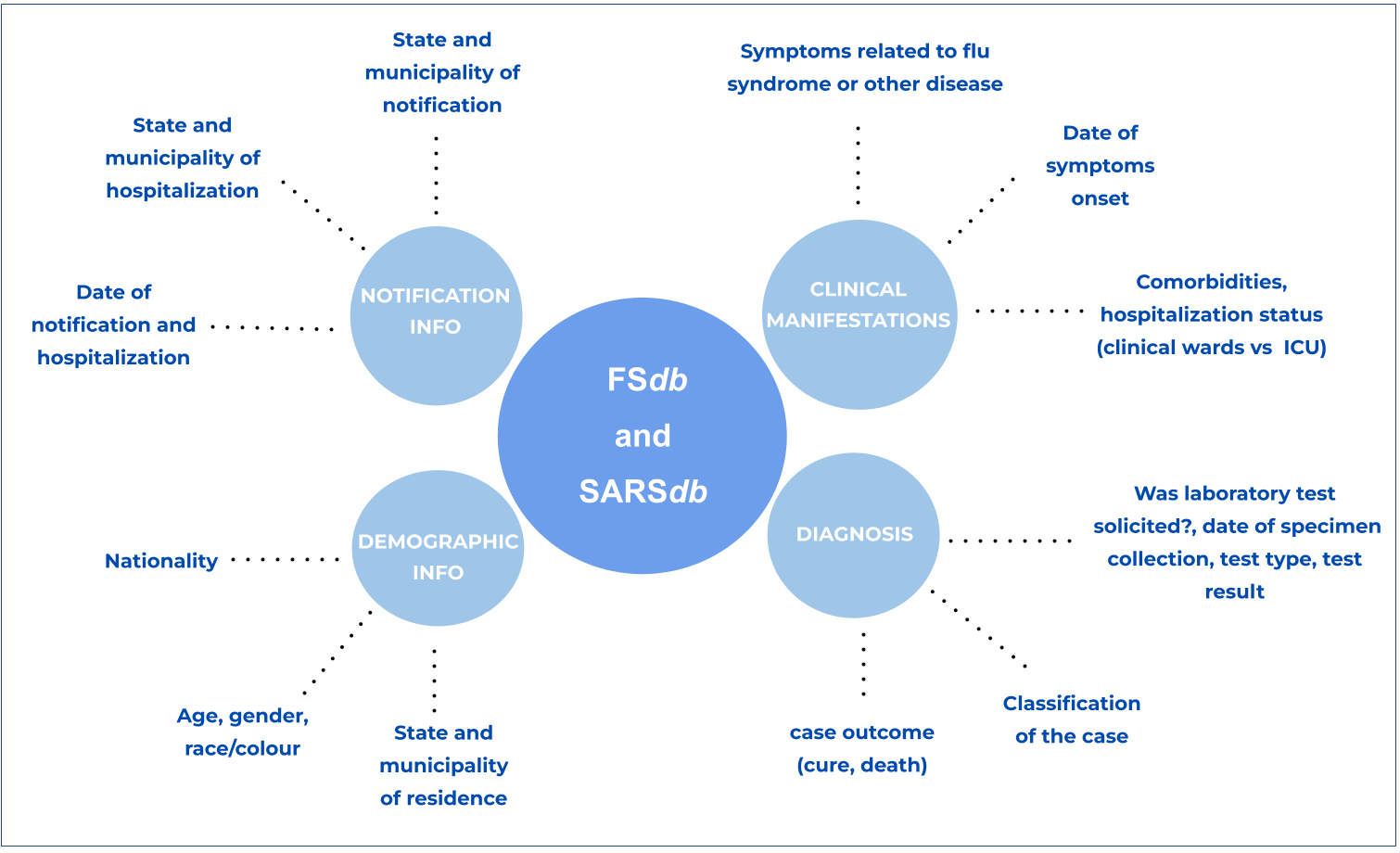}
\caption{Schematic diagram of main variables in FS\textit{db} and SARS\textit{db}.}
\label{fig2}
\end{figure}

FS\textit{db} and SARS\textit{db} contain information provided by private and public health institutions, which report suspected and confirmed cases via the e-SUS NOTIFICA system. FS\textit{db} is organized according to each federal unit (i.e., state) of the Brazilian federation, while SARS\textit{db} files are organized annually (one per year). FS\textit{db} is available in .csv format, with a total of 30 variables and a size of around 15 GB in December 2021. In turn, SARS\textit{db}, also available in .csv format, contains 161 variables and a size of around 2 GB in December 2021. Variables provide spatial and temporal information on reported cases, as well as patient clinical (symptoms, comorbidities, etc.) and demographic (age, municipality of residence, etc.) data (Figure \ref{fig2}). Additionally, SARS\textit{db} also contains each patient’s vaccination status against influenza or COVID-19 (vaccination cycle, type of vaccine, etc.) and hospitalization information (clinical ward or intensive care unit). FS\textit{db} and SARS\textit{db} are updated weekly with newly included case information.

\begin{figure}[H]
\centering
\includegraphics[height=9cm,width=17cm]{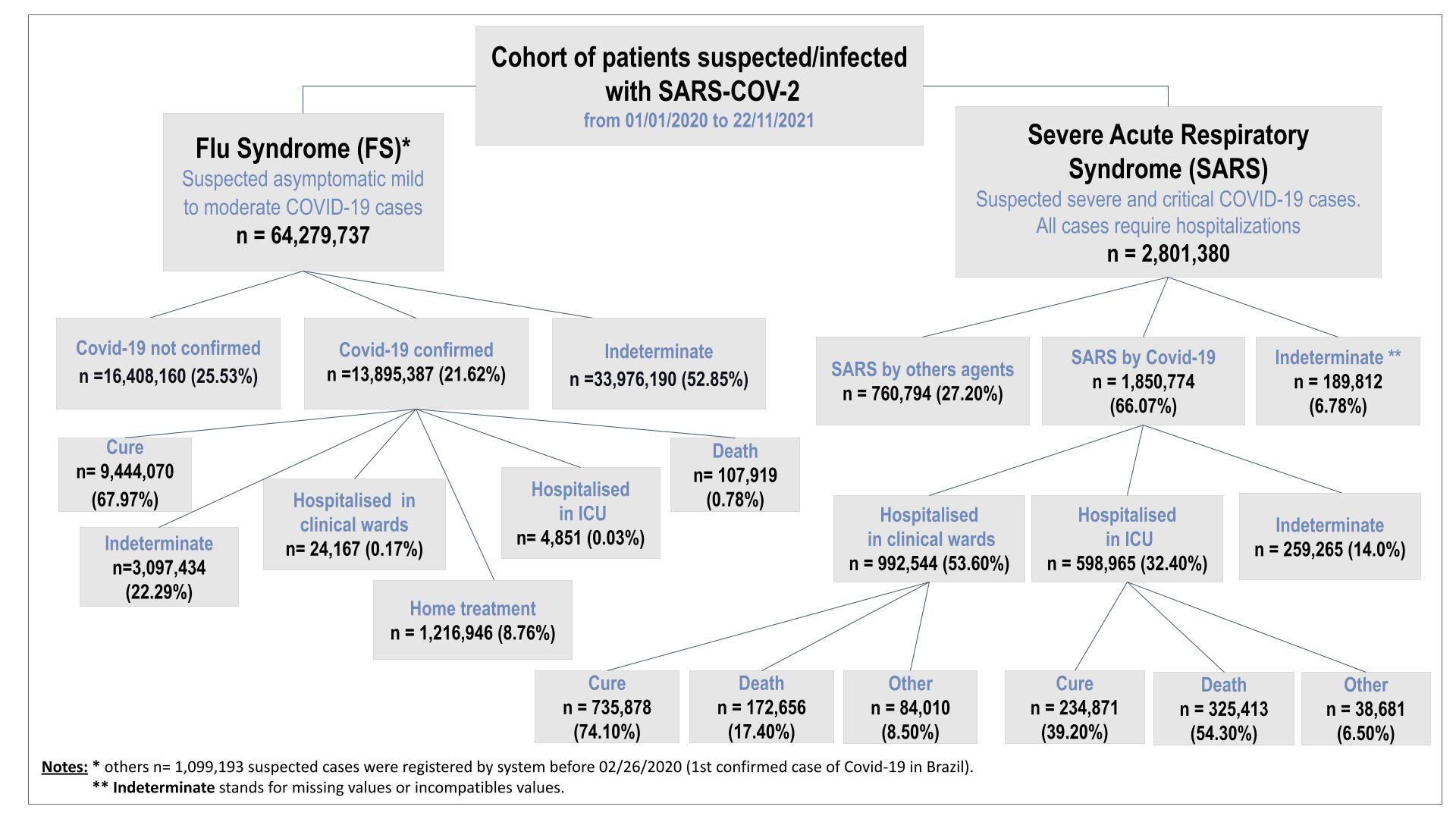} 
\caption{Cases reported in FS\textit{db} and SARS\textit{db} with corresponding final diagnostic classification, hospitalization status and clinical outcome information.}
\label{fig3}
\end{figure}

As of November 2021, more than 21\% of the cases registered in FS\textit{db} were confirmed as COVID-19 (13,895,387 registries), with more than 67\% resulting in cure and less than 1\% in deaths (Figure \ref{fig3}). By contrast, SARS\textit{db} contained reports of more than two million suspected severe-to-critical cases. Of these, 66.1\% were confirmed to be COVID-19, with 32.4\% being admitted to an ICU and a resulting death rate >50\%. Among patients who were hospitalized but not admitted to ICU, 17.5\% died from COVID-19, and more than 74\% achieved cure (Figure \ref{fig3}).

Assigning a definitive classification to cases of suspected SARS-CoV-2 infection remains a challenge for Brazilian health authorities, as evidenced by high percentages of indeterminate cases in FS\textit{db} and SARS\textit{db} (Figure \ref{fig3}). As is typical for other diseases in Brazil (such as arboviruses and leprosy)\cite{oliveira2020interdependence,de2021estimating}, many cases go undetected, and those reported in health databases are usually not updated with a final classification due to the huge number of registered patients and lack of laboratory facilities and testing. This limitation affects the estimation of true disease prevalence and the implementation of appropriate control measures. Under this scenario, mathematical, statistical and computational methodologies have been extensively employed to better understand the reality reflected by the information contained in Brazilian databases to address deficiencies, and provide more reliable estimates of true disease prevalence by inferring diagnosis status for cases not definitively classified in the database records\cite{oliveira2021mathematical,de2021estimating,veiga2021classification}.

\medskip

\subsubsection*{3. Vaccination database - Vac\textit{db}} % correção ingles concluída

\medskip

The Brazilian national vaccination campaign against COVID-19 started in January 2021. The National Vaccination Operational Plan (NVOP)\cite{pno}, elaborated by the MoH, laid out the strategy necessary to vaccinate Brazil’s population as quickly as possible. Vaccination began incrementally based on priority groups as defined by the NVOP. All vaccines introduced by the Brazilian Unified Health System underwent technical evaluations and were approved by the Brazilian Health Regulatory Agency (Anvisa) for emergency use. Currently, AZD1222 Vaxzevria (Astrazeneca/Oxford/Fiocruz), COVID-19 Vaccine (Vero Cell), Inactivated/Coronavac (Sinovac/Butantan), BNT162b2/COMIRNATY Tozinameran (INN) (Pfizer/BioNTech/Wyeth) and Ad26.COV2.S (Janssen/Johnson \& Johnson) vaccines are included in the National Immunization Program (PNI)\cite{who2020,pno}. 

All data related to the vaccination campaign are made available by the MoH through the National Immunization Program’s information system. The Opendata SUS Vac\textit{db} provides information on the doses administered (vaccine type, patient vaccination status) as well as patient demographic characteristics. Additional information related to administrative codes (system patient identifiers, IBGE or vaccine manufacturer) is also present. The data set, available in .csv format, contains 30 variables, is updated weekly, and, as of December 2021, had a total size around 80 GB. Table \ref{tab:vacina} lists the main variables contained in Vac\textit{db}.

\begin{table}[htbp]
\caption{\label{tab:vacina}Description of main variables in Vac\textit{db}.}
\begin{tabular}{|l|l|l|l|}
\hline
\multicolumn{1}{|c|}{\textbf{Field name (in original db)}} & \multicolumn{1}{c|}{\textbf{Field label}}&  \multicolumn{1}{c|}{\textbf{Type}}& \multicolumn{1}{c|}{\textbf{Description}} \\ \hline
paciente\_id                                  & Identifier               & String        & A randomly generated unique identifier for the patient. \\ \hline
paciente\_idade                               & Age                      & Integer       & Patient age                                             \\ \hline
paciente\_dataNascimento                      & Birth date               & Date          & Patient’s date of birth                                   \\ \hline
paciente\_enumSexoBiologico                   & Sex                      & String        & Biological sex of the patient                           \\ \hline
paciente\_racaCor\_valor                      & Self-reported skin color & String      &  Patient’s race/skin color                                     \\ \hline
paciente\_endereco\_nmMunicipio               & Municipality of residence & String  & Patient’s municipality of residence                            \\ \hline
paciente\_endereco\_uf                        & State of residence       & String        & Patient's state of residence                            \\ \hline
vacina\_nome                                  & Vaccine name             & String        & Vaccine name                                            \\ \hline
vacina\_dataAplicacao                         & Vaccine application date & Date          & Vaccine application date                                \\ \hline
vacina\_descricao\_dose                       & Dose description   & String        & Description of the dose (1st, 2nd, booster, etc.)             
\\ \hline
\end{tabular}
\end{table}

From the start of the NVOP campaign until November 22, 2021, 154,007,191 people in the country had received an initial dose, representing 72.2\% of the population, while 58.1\% (123,912,270) of the population had a complete vaccination schedule (2nd dose or single dose), and approximately 6.5\% had received a booster dose.

\subsection*{Aggregated data}
\medskip

\subsubsection*{1. Human Mobility}
\medskip

Google has provided a Community Mobility Report\cite{google} to assess how human mobility has been affected by the spread of COVID-19 since February 2020\cite{jorge2021assessing,oliveira2021mathematical,kerr2020covid}. The report compares the median attendance of people at a given location and day to the median attendance value at the same location for that day of the week, calculated for the pre-pandemic period from January 3 through February 6, 2020 (these values are defined as a baseline). The resulting metric, available for each country and regions within, reflects the percentage change in the number of people who visited grocery stores or pharmacies, parks, transit stations, retail and recreation entities, residences, and workplaces compared to the same period defined before. Google’s mobility database contains 15 variables and is updated daily. The information available is based on a sample of users who have allowed the company to track their location history, which may introduce bias depending on the analysis performed.

\begin{figure}[htbp]
\centering
\includegraphics[width=0.63\textwidth]{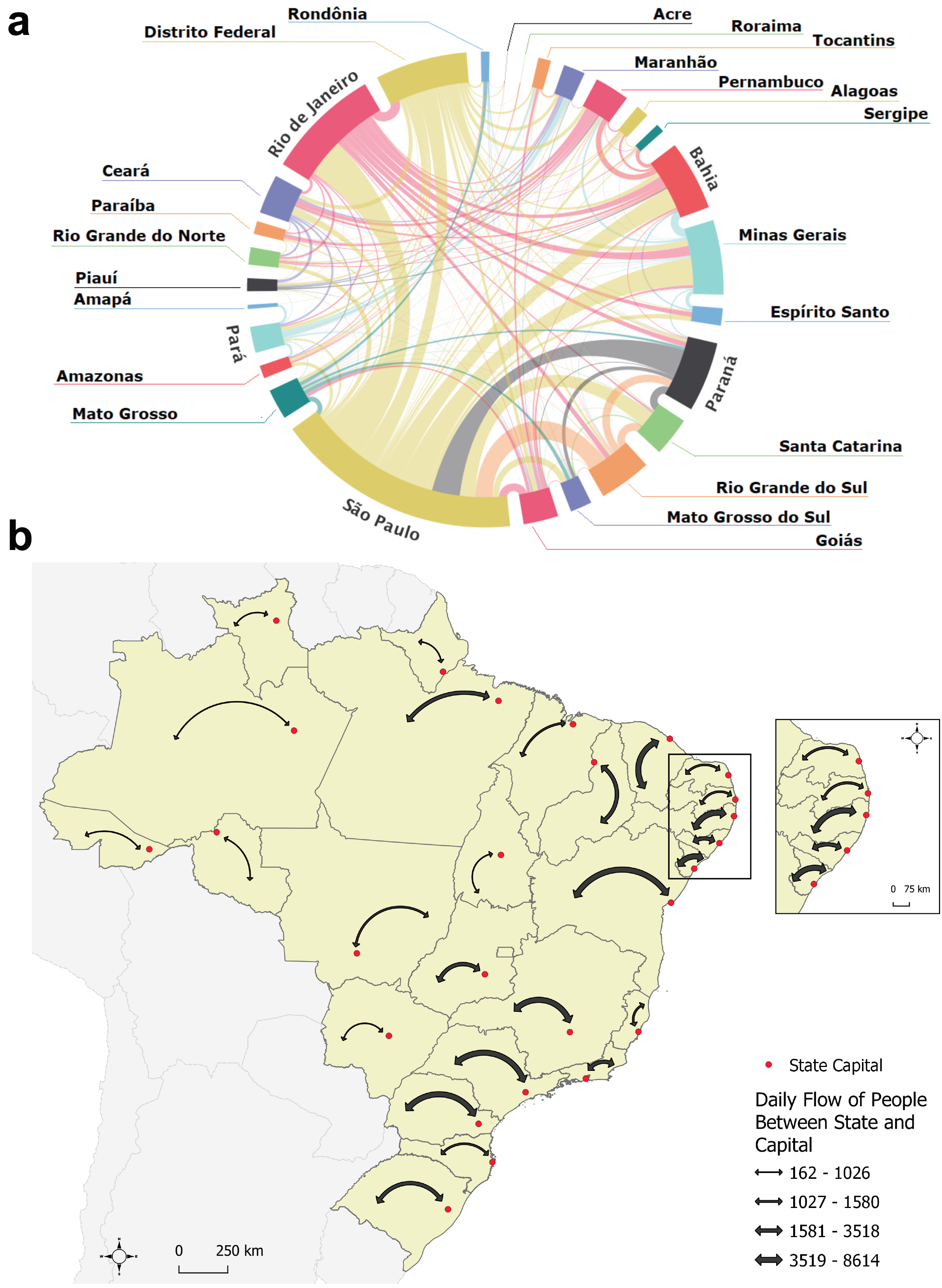}
\caption{a) Illustration of annual traffic flows among Brazilian states via air/road/river networks (2010 data). State and flow colors are for identification purposes only. b) Estimation of average daily traffic between state capitals (shown as red dots) and inland cities in each Brazilian federal unit (2016 data).}
\label{mob}
\end{figure}

In addition, IBGE provides information on the flow of people between Brazilian cities on roads, rivers and via air. Air mobility data indicates the average number of people travelling between the country’s airports\cite{ibge2019} for the years 2010, 2019 and 2020. Another report contains data on intercity vehicle (bus) transport via roads or rivers estimated for 2016\cite{ibge2016}. Figure \ref{mob} illustrates the Brazilian interstate and intercity flow network (via air, road and river).

It is important to note that the IBGE databases do not indicate the direction of flow. Therefore, it is impossible to discern whether planes/vehicles are arriving or leaving a given city by air/road/river. Moreover, the road/river mobility data only considers the average number of vehicles per week, and does not specify vehicle occupancy rates.
 
\medskip

\subsubsection*{2. Population inequality} 
\medskip

The Brazilian Deprivation Index (BDI)\cite{allik2020developing,ibp} is a composite measure of health and social vulnerability, and represents the most up-to-date area-based measures of deprivation in Brazil, allowing to estimate socioeconomic inequalities at multiple scales ranging from census tracts, municipalities, states, macrorregions, or the entire country. The BDI combines information on the income, education and household conditions in a given region to measure social inequalities. The index combines the z-scores of three factors: the percentage of families with per capita incomes less than half the minimum wage (~US\$117); and the percentage of illiterates older than seven years; the percentage of people without adequate access to potable water, sanitation, garbage disposal, bathroom or shower.

Limitations of using this metric exist, since it comprises three dimensions of deprivation, excluding others such as employment, crime, health, education, and access to public services. Comparisons over time are also not possible since the data that compose the index were gathered from the 2010 Brazilian census. Because of this, the measure does not need to be frequently updated on our data lake. Still, the BDI contributes to essential insights on characterising the impact of the COVID-19 spread on the Brazilian population\cite{pereira4081979profile}.

The BDI database we collect measures deprivation at the municipal level. It has 15 variables, 5,566 registers (rows) and is currently 734 Kb in size. The data is shared under a Creative Commons Share-Alike license, with copyright owned by the author's institutions (CIDACS/Fiocruz-Bahia and the University of Glasgow). The database, along with its extensive documentation, can be accessed at \url{https://researchdata.gla.ac.uk/980/} or \url{https://cidacs.bahia.fiocruz.br/ibp/painel-en/}.

\medskip

\subsubsection*{3. Non-pharmaceutical interventions}
\medskip

Like most countries, as the epidemic unfolded the Brazilian government began to enact non-pharmaceutical interventions (NPIs) to mitigate the effects of COVID-19, specially the surge in hospitalization of COVID-19 patients. Some of these NPI were lifted over time following declines in number of cases and deaths between each infection wave. Several studies reported the effects of NPIs on the incidence of COVID-19\cite{jorge2021assessing,oliveira2021mathematical,kerr2020covid}, highlighting how, coupled with human mobility patterns, they can be applied to gauge the intensity of government stringency needed to achieve control of viral transmission rates\cite{pataro2021control}. The governmental decrees that have been issued during this period can be extracted from state- and federal-level Official Bulletins (in Portuguese, Diário Oficial). Although the federal-level bulletin can be bulk downloaded in computer-readable XML format (\url{https://www.in.gov.br/acesso-a-informacao/dados-abertos/base-de-dados}), the same is not possible for state-level bulletins, which are decentralized and usually do not permit automatic data query or retrieval, which also occurs with municipality-level decrees.

As a result of a collaboration agreement with legal information start-up JusBrasil (\url{http://jusbrasil.com.br}), we accessed state and city level text files describing NPIs enacted during COVID-19 issued by governments through JusBrasil's proprietary Application Programming Interface, allowing fast retrieval of COVID-19-related legislation during the period. The automatic retrieval was followed by manual human curation, when each decree was read, categorized, and multiple meta-data were systematically derived, including the geographic coverage of the measures and the validity period of the decree. Based on the type of measure processed, a metric was created, defined as the stringency index \cite{jorge2021assessing}, which combines the different enacted laws to summarize the level of governmental strictness over any given period of the pandemic. The contents of evaluated decrees (in Portuguese) and the resulting stringency index over time are freely available and regularly updated within the PAMepi platform at \url{https://pamepi.rondonia.fiocruz.br/en/decree_en.html}.

\subsection*{Data stream limitations} 

In an attempt to enhance our understanding of the spread of respiratory disease in Brazil, particularly the COVID-19 pandemic, our data lake incorporates multiple databases that can be accessed, downloaded and routinely updated via automated scripts. However, as changes in disease status occur due to human interventions (e.g., changes in mobility patterns, non-pharmaceutical interventions, vaccine campaigns), the source databases may add or remove variables (or even include additional categories within variables). For example, this was observed in SARS\textit{db} and Vac\textit{db} when new variables were added to account for individuals infected with SARS-CoV-2 who were vaccinated, as well as to incorporate the administration of additional doses during the ongoing national vaccination campaign, respectively. Therefore, regular monitoring of data sources, which we perform, is necessary to update the automated script used to execute data lake updates.

\medskip

\subsection*{Relationships between files}
\medskip

We created datasets containing community-level aggregated information extracted from the variables in the source databases. The integration of variables from different sources facilitates data analysis and epidemiological study, since the resulting aligned dataset contains relevant datapoints collected since the beginning of the pandemic in Brazil, as well as prior to the pandemic. For example, users can analyze a daily time series of mild-to-moderate cases resulting from the FS\textit{db}, or hospital occupancy and death rates from the SARS\textit{db}, in addition to daily vaccine doses administered, among other variables. The dataset enables real-time assessment of the pandemic in each Brazilian municipality through the lens of data visualization, which is essential to fostering community involvement during local and world health challenges\cite{johns,whodash21,worldometer21,PAMEpi2020}. Finally, the dataset metadata we provide also facilitates federated data access and analytics, which are important for future integrative research and preparedness for the next pandemic.

Users can access the updated dataset using a data explorer and dictionary, which are available at \url{https://pamepi.rondonia.fiocruz.br/en/aggregated_en.html}. In addition, all codes used to create the dataset are freely available from our GitHub repository\cite{PAMEpi2020}. 

\section*{Technical Validation}

After performing pre-processing and data curation, the databases directly related to COVID-19 information (FS\textit{db}, SARS\textit{db} and Vac\textit{db}) were verified through exploratory analyses to ascertain the technical quality of the databases. Verification consisted of evaluating the completeness of the main variables contained in these databases, while checking for inconsistencies in date-type variables. To assess the completeness of a given variable, we constructed a new column termed ‘indeterminate’ that served to quantify the amount of missing and incompatible data. Accordingly, database completeness is represented by the percentage of completion of variables, varying from poorly filled (0\%) to well-filled (100\%).

Completeness was evaluated by selecting key variables from each database. More extensive descriptions of completeness are provided for each individualised database as Supplementary material. These reports include initial exploratory analyses in both table and plot format. The period considered for this evaluation was between January 1, 2020 and November 22, 2021.

FS\textit{db} and SARS\textit{db} do not contain individual identification codes. Therefore, no procedure to identify duplicate records was carried out. We then labelled categorical variables and categorised age variables into groups to evaluate completeness. The spatial unit used for this analysis was the state in which cases were notified. The completeness analysis in FS\textit{db} examined the following demographic and clinical variables: sex, age group, final classification, case evolution, symptoms, test result and test type. For SARS\textit{db}, we assessed the completeness of sex, age group, scholarity, residential area (urban or rural), date of hospitalisation, ICU admission, case evolution, final classification and symptoms (each is represented by a variable in SARS\textit{db}), such as fever, cough, throat pain, dyspnea, respiratory discomfort, saturation, diarrhea, vomiting, abdominal pain and fatigue. We also explored the completeness of the symptoms variable using cross-tabulations of final classification, case evolution, mortality and COVID-19 case distribution by age group.

Vac\textit{db} contains an identification code for each individual, as one person can receive more than one vaccine dose. Considering the data available to date, the maximum number of duplicate records for each subject should be three (first/single dose, second and third/booster dose). Therefore, we normalized Vac\textit{db} to identify each individual in a row versus the type of dose applied, as the original database is organized\cite{PAMEpi2020}. We then evaluated the completeness of first dose, second dose, third dose/reinforce, single-dose and single-dose/reinforce variables across Brazil, and in each state by age and ethnicity.

Table \ref{tab:completude} details completeness information for FS\textit{db}, SARS\textit{db} and Vac\textit{db} at a national level. Age group, sex and symptoms presented the highest completeness in FS\textit{db} (>99\%), while final classification and evolution were lower (47.14\% and 33.34\%, respectively). For SARS\textit{db}, 63.88\% of the records contained indeterminate information on scholarity. The lowest percentage of completeness was found for the sex variable. Among symptoms, the variables dyspnea and cough showed higher completeness, while fatigue and abdominal pain were lower (Table \ref{tab:completude}). Completeness for all vaccination dose types in Vac\textit{db} by age group reached over 99\% (Table \ref{tab:completude}). Sex by all vaccination dose types achieved almost 100\% completeness. Ethnicity by all vaccination dose types attained around 75\% completeness (see Completeness data analyses available at \url{https://pamepi.rondonia.fiocruz.br/en/data_en.html}).

\begin{table}[htbp]
\caption{\label{tab:completude}Completeness information for FS\textit{db}, SARS\textit{db} and Vac\textit{db} at national level.}
\begin{center}
\begin{tabular}{lcc}
\hline
\multicolumn{3}{l}{\textbf{FS\it db}}                                                                                                       \\ \hline
\multicolumn{1}{c|}{Demographic and health condition} & \multicolumn{1}{c|}{Indeterminate$^{\ast}$ (\%)} & \multicolumn{1}{c}{Completeness (\%)} \\ \hline
\multicolumn{1}{l|}{Sex}                        & \multicolumn{1}{c|}{0.07}                      & \multicolumn{1}{c}{99.93}             \\
\multicolumn{1}{l|}{Age group}                     & \multicolumn{1}{c|}{0.04}                      & \multicolumn{1}{c}{99.96}             \\
\multicolumn{1}{l|}{Symptoms}                      & \multicolumn{1}{c|}{0.13}                      & \multicolumn{1}{c}{98.87}             \\
\multicolumn{1}{l|}{Case evolution}                & \multicolumn{1}{c|}{66.66}                     & \multicolumn{1}{c}{33.34}             \\
\multicolumn{1}{l|}{Final classification}          & \multicolumn{1}{c|}{52.86}                      & \multicolumn{1}{c}{47.14}             \\
\multicolumn{1}{l|}{Test type}                     & \multicolumn{1}{c|}{14.83}                      & \multicolumn{1}{c}{85.17}             \\
\multicolumn{1}{l|}{Test result}                   & \multicolumn{1}{c|}{27.46}                      & \multicolumn{1}{c}{72.54}             \\ \hline
\multicolumn{3}{l}{\textbf{SARS\it db}}                                                                                                       \\ \hline
\multicolumn{1}{c|}{Demographic and health condition} & \multicolumn{1}{c|}{Indeterminate(\%)} & \multicolumn{1}{c}{Completeness (\%)} \\ \hline
\multicolumn{1}{l|}{Sex}                        & \multicolumn{1}{c|}{0.03}                      & \multicolumn{1}{c}{99.97}             \\
\multicolumn{1}{l|}{Age group}                     & \multicolumn{1}{c|}{0.12}                      & \multicolumn{1}{c}{99.88}             \\
\multicolumn{1}{l|}{Residential area (urban or rural)}                & \multicolumn{1}{c|}{7.18}                      & \multicolumn{1}{c}{92.82}             \\
\multicolumn{1}{l|}{Scholarity}                    & \multicolumn{1}{c|}{63.88}                     & \multicolumn{1}{c}{36.12}             \\
\multicolumn{1}{l|}{Case evolution}                & \multicolumn{1}{c|}{12.02}                     & \multicolumn{1}{c}{87.98}            \\
\multicolumn{1}{l|}{Final classification}          & \multicolumn{1}{c|}{6.52}                      & \multicolumn{1}{c}{93.48}             \\
\multicolumn{1}{l|}{Date of hospitalization}       & \multicolumn{1}{c|}{16.06}                     & \multicolumn{1}{c}{83.94}            \\
\multicolumn{1}{l|}{ICU admission}        & \multicolumn{1}{c|}{12.12}                     & \multicolumn{1}{c}{87.88}             \\
\multicolumn{1}{l|}{Dyspnea}                       & \multicolumn{1}{c|}{13.23}                     & \multicolumn{1}{c}{86.77}             \\
\multicolumn{1}{l|}{Cough}                         & \multicolumn{1}{c|}{13.80}                     & \multicolumn{1}{c}{86.20}             \\
\multicolumn{1}{l|}{Fever}                         & \multicolumn{1}{c|}{17.42}                     & \multicolumn{1}{c}{82.58}            \\
\multicolumn{1}{l|}{Throat pain}                   & \multicolumn{1}{c|}{30.78}                     & \multicolumn{1}{c}{69.22}             \\
\multicolumn{1}{l|}{Diarrhea}                      & \multicolumn{1}{c|}{31.75}                     & \multicolumn{1}{c}{68.25}             \\
\multicolumn{1}{l|}{Vomit}                         & \multicolumn{1}{c|}{32.86}                     & \multicolumn{1}{c}{67.14}             \\
\multicolumn{1}{l|}{Fatigue}                       & \multicolumn{1}{c|}{42.62}                     & \multicolumn{1}{c}{57.38}             \\
\multicolumn{1}{l|}{Abdominal Pain}                & \multicolumn{1}{c|}{45.79}                     & \multicolumn{1}{c}{54.21}             \\
\multicolumn{1}{l|}{Respiratory discomfort}        & \multicolumn{1}{c|}{19.86}                     & \multicolumn{1}{c}{80.14}             \\
\multicolumn{1}{l|}{Saturation}                    & \multicolumn{1}{c|}{17.72}                     & \multicolumn{1}{c}{82.28}             \\ \hline
\multicolumn{3}{l}{\textbf{Vac\it db}}   \\ \hline
\multicolumn{1}{l|}{Vaccine status} & \multicolumn{1}{c|}{Indeterminate(\%)} & \multicolumn{1}{c}{Completeness (\%)} \\ \hline
\multicolumn{1}{l|}{Single dose}                  & \multicolumn{1}{c|}{0.00}                      & \multicolumn{1}{c}{100.00}             \\
\multicolumn{1}{l|}{First dose}                   & \multicolumn{1}{c|}{0.01}                      & \multicolumn{1}{c}{99.99}             \\
\multicolumn{1}{l|}{Second dose}                  & \multicolumn{1}{c|}{0.01}                      & \multicolumn{1}{c}{99.99}             \\
\multicolumn{1}{l|}{Third Dose/Reinforce}         & \multicolumn{1}{c|}{0.05}                      & \multicolumn{1}{c}{99.95}             \\
\multicolumn{1}{l|}{Single Dose/Reinforce}        & \multicolumn{1}{c|}{0.09}                      & \multicolumn{1}{c}{99.91}             \\\hline
\multicolumn{3}{l}{$^{\ast}${\small stands for missing values or incompatible values.}} 
\end{tabular}
\end{center}
\end{table}

The validation presented in the material available at \url{https://pamepi.rondonia.fiocruz.br/en/data_en.html} shows that the variable completeness varies in accordance with the state in which a case was notified. This may serve to indicate quality regarding data collection processes on a regional health district level (data is then forwarded to the MoH data systems). However, an in-depth analysis of the quality of data collection is outside the scope of this manuscript.

\section*{Usage Notes} % correção ingles concluída

The COVID-19 pandemic continues to pose a threat to global health. Populations worldwide face severe inequalities with regard to the distribution of resources essential to diagnosing, treating and preventing COVID-19\cite{pereira4081979profile}. Concomitantly, extreme poverty and its consequences, particularly poverty-related infectious diseases, such as tuberculosis, HIV/AIDS, malaria, etc., persistently imperil the health of vulnerable populations, and scant attention has been paid to these conditions due to limited health services during the pandemic.  Therefore, the data lake we have constructed represents a powerful resource for studying past and current aspects of the COVID-19 pandemic, as well as its impact on other health, economic and social outcomes.

We have designed the PAMEpi data architecture to incorporate different data sources in order to enhance the power of a data-driven approach to understanding the COVID-19 pandemic in Brazil. To enrich the available health data, we aggregated data on human mobility, social inequality and non-pharmaceutical government interventions to better reflect the complexities of the COVID-19 pandemic. We have provided a description of all routines employed to collect, store, organise, integrate and use data in real-time for potential use in exploratory analysis, modelling and visualisation at both individual and aggregated levels. Furthermore, this approach is not solely limited to COVID-19, as the health data ingested also contains reports on other respiratory diseases circulating in the country. Therefore, our efforts can be extended to other diseases (particularly respiratory diseases) to include additional important variables (e.g., biogeoclimatic parameters) and potentially aid in the response to future emerging or re-emerging infectious diseases.% detection of novel pathogens.

\section*{Code availability}

As described in the Methods section, all coding has been documented and is freely available in R or Python languages on the project’s Github page (\url{https://github.com/PAMepi}). 

\bibliography{main} 

\section*{Acknowledgements}
This study was financed by Bill and Melinda Gates Foundation and Minderoo Foundation HDR UK, through the Grand Challenges ICODA COVID-19 Data Science, with reference number 2021.0097 and the Fiocruz Innovation Promotion Program - Innovative ideas and products - COVID-19, orders and strategies INOVA-FIOCRUZ, with reference Number VPPIS-005-FIO-20-2-40. We thank Andris K Walter for English revision and manuscript proofing and suggestions.

\section*{Author contributions statement} 

N.B.S., F.M.H.S.F., L.I.O.V., A.C.S.F., F.A.C.P. and J.F.O. contributed to collecting and curating the database. N.B.S., L.I.O.V., G.L.O. and J.F.O. contributed to technical validation on the databases. All authors contributed to contributed to analyse the results. N.B.S and J.F.O. contributed to writing the manuscript. All authors contributed to and reviewed the final submitted manuscript. 

\section*{Competing interests}

The authors declare no competing interests.

\end{document}